\pdfoutput=1
\documentclass[amsmath,amssymb,aps,prb,superscriptaddress,twocolumn,floatfix]{revtex4-2}
\usepackage[normalem]{ulem}
\usepackage{graphicx} % Include figure files
\usepackage{wrapfig}
\usepackage{bm}
\usepackage[usenames]{color}
\usepackage{MnSymbol}
\usepackage{hyperref} 
\usepackage{soul}
\usepackage{xcolor}
\definecolor{blue}{HTML}{3f76b5}
\definecolor{red}{HTML}{ba1616}
\usepackage{subfigure}

\newcommand{\spincor}{\langle \mathbf{S}_i \cdot \mathbf{S}_j \rangle}

\newcommand{\para}{}
\begin{document}
\title{Quantum Melting of Generalized Wigner Crystals in Transition Metal Dichalcogenide Moir{\'e} Systems}
\author{Yiqing Zhou}
\affiliation{Laboratory of Atomic and Solid State Physics, Cornell University, Ithaca, NY 14853, USA}
\author{D. N. Sheng}
\affiliation{Department of Physics and Astronomy, California State University, Northridge, CA 91330, USA}
\author{Eun-Ah Kim}
\affiliation{Laboratory of Atomic and Solid State Physics, Cornell University, Ithaca, NY 14853, USA}
\affiliation{Department of Physics, Ewha Womans University, Seoul, South Korea}

\begin{abstract}
Generalized Wigner Crystal (GWC) is a novel quantum phase of matter driven by further-range interaction at fractional fillings of a lattice. The role of further range interaction as the driver for the incompressible state is akin to Wigner crystal. On the other hand, the significant role of commensurate filling is akin to the Mott insulator. Recent progress in simulator platforms presents unprecedented opportunities to investigate quantum melting in the strongly interacting regime through synergy between theory and experiments. However, the earlier theory literature presents diverging predictions. We study the quantum freezing of GWC
through large-scale density matrix renormalization group simulations of a triangular lattice extended Hubbard model. We find a single first-order phase transition between the Fermi liquid and the $\sqrt{3}\times\sqrt{3}$ GWC state. The GWC state shows long-range antiferromagnetic $120^\circ$ N{\'e}el order. Our results present the simplest answers to the question of the quantum phase transition into the GWC phase and the properties of the GWC phase. 
% {\color{blue} With the increasing availability of GWC quantum simulation platforms, our results offer critical insights on quantum freezing and inform future experiments.} 

% {\color{red} Moreover, exponentially decaying spin correlation imply gap to spin excitations, ruling out gapless quantum spin liquids.} 
\end{abstract}
\maketitle

\para Mott insulator \cite{mott1949basis} and Wigner crystal \cite{wigner1934interaction} are two distinct long-standing theoretical mechanisms for interaction-driven
 metal-insulator transitions. 
 Mott insulator is driven by on-site repulsion on a lattice when the filling is one electron per site.  On the other hand, Wigner crystal (WC) is driven by long-range Coulomb interaction in a continuum system.  
 In a WC, the crystalline lattice emerges due to long-range interaction~\cite{sung2023observation}.
However, the advancement of simulator platforms ~\cite{xu2020correlated, Li2021Naturec,norcia2021two,bland2022two,su2023dipolar} on artificially defined lattice put forth a new paradigm that combines the two mechanisms: generalized Wigner crystals (GWC).  The GWC occurs when a lattice is partially filled with charge carrier under further range interactions~\cite{xu2020correlated, Li2021Naturec, padhi2021generalized,tan2023doping}. As in Wigner crystals, long-range interaction is critical for the formation of GWC. However, due to the lattice potential,  different charge order forms at different commensurate fillings.  While the strong coupling limit of GWC can be readily established using classical considerations \cite{Matty2022NatCommun}, quantum melting of GWC remains a theoretically challenging problem.  However, rapid developments in diverse settings for simulating GWC imply unprecedented opportunity to understand quantum melting through synergy between theory and experiments~\cite{xu2020correlated, Li2021Naturec,norcia2021two,bland2022two,su2023dipolar}.
 
\begin{figure*}[]
\includegraphics[width=1.9\columnwidth]{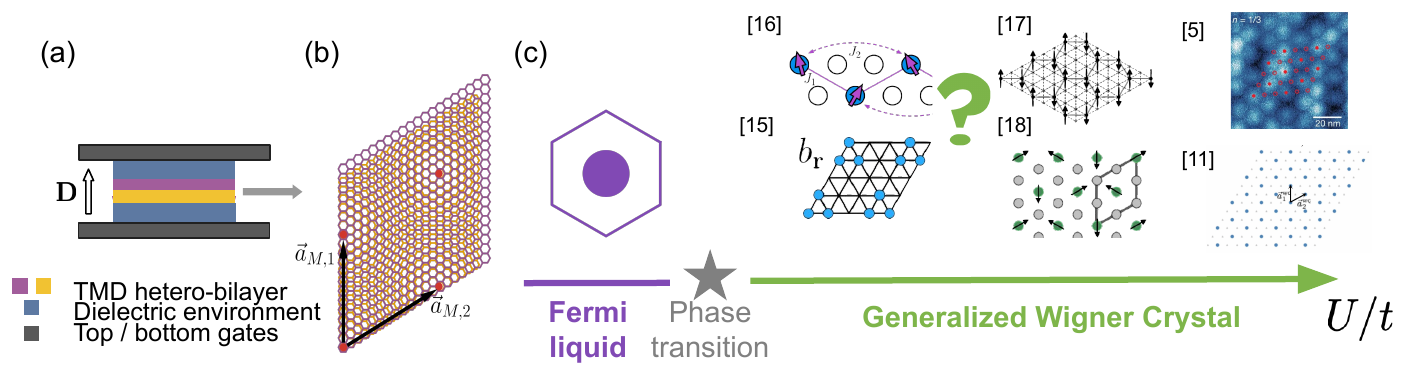}
  \caption{ (a) A side-view schematic of the experimental setup. The sample comprises the TMD hetero-bilayers (purple and yellow rectangles) encapsulated in a dielectric environment (blue rectangles on two sides).  The sample is placed between two metallic gates (dark grey rectangles). Symmetric gating controls the carrier density $\nu$ in the system, while asymmetric gating tunes the strength of the external displacement field $\mathbf{D}$.
  (b) An illustration of the moir{\'e} pattern formed by hetero-bilayer TMDs. Each monolayer forms a honeycomb lattice. The two layers, with an $8\%$ lattice constant mismatch, are stacked with angle aligned. The red dots mark the emergent moir{\'e} superlattice sites, which form a triangular lattice. The primitive vectors for the moir{\'e} superlattice are marked by black arrows. 
  (c) Problem statement. In the small $U/t$ limit, Fermi liquid (FL) is expected. In flat band (large $U/t$) limit, $\sqrt{3} \times \sqrt{3}$ charge order is observed by STM experiment. The nature of the phase transition between the two limits remains an open question. A number of recent studies, as shown in the middle of the figure, predict the order of the transition and the charge order pattern near the transition differently. 
}
  \label{fig:model}
\end{figure*}

\para Of particular interest is experimental platforms based on  hetero-transition metal dichalcogenide (TMD) bilayers such as AA-stacked MoTe$_2$/WSe$_2$,  which forms a \textit{triangular} moir{\'e} superlattice due to lattice constant mismatch (see Fig.\ref{fig:model}(b)). 
 Unlike twisted graphene-based moir{\'e} systems,  these hetero-TMD bilayers can be accurately modeled by a single isolated band of fermions living on a triangular lattice~\cite{Wu2018Phys.Rev.Lett.a} at low energies.  The type II band alignment with a large band offset freezes the layer degree of freedom~\cite{Zhang20162DMater.} and the spin-valley locking freezes the spin degree of freedom~\cite{Xiao2012Phys.Rev.Lett.a}.  Hence single band extended Hubbard model that includes further range interaction can faithfully capture the system.  Moreover,  the strong coupling limit of the GWC at various commensurate fillings is well-established~\cite{Matty2022NatCommun,Li2021Naturec}. 
To study the quantum melting of the GWC however,  one needs an approach that can handle the charge fluctuation and strong interaction in a sufficiently large system size. 

\para Interestingly, predictions from low-energy effective theory in the weak coupling limit \cite{Musser2022Phys.Rev.Bc,Musser2022Phys.Rev.Bb}, those from Hartree-Fock (HF) studies~\cite{Pan2022Phys.Rev.Ba}, and those from momentum space exact diagonalization (ED)~\cite{Morales-Duran2022a} all diverge from each other, as shown in 
Fig. \ref{fig:model}(c). 
While Ref.~\cite{Musser2022Phys.Rev.Bb,Musser2022Phys.Rev.Bc} showed a direct continuous transition between FL and GWC insulator with spinon Fermi surface is possible, the resulting GWC pattern is distinct from strong coupling limit predictions \cite{Matty2022NatCommun} and scanning tunneling microscopy in the flat band limit \cite{Li2021Naturec}, the $\sqrt{3}\times\sqrt{3}$ phase.
On the other hand, the HF \cite{Pan2022Phys.Rev.Ba}  and the ED \cite{Morales-Duran2022a}  both found a first order transition into the GWC with spin ordering albeit with different spin orders. 
In this paper, we model a hetero-TMD bilayer by a single-band extended Hubbard model on a triangular lattice with valley degeneracy,  and study the metal-insulator transition at 1/3 filling through a large-scale density matrix renormalization group (DMRG).  
 
\begin{figure}[h]
\includegraphics[width=1.0\columnwidth]{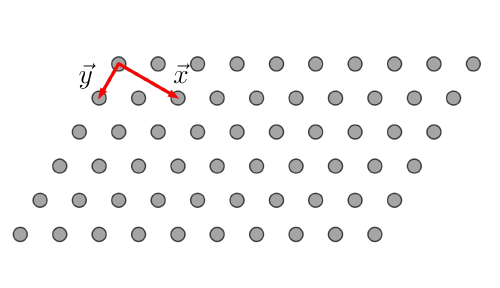}
  \caption{A schematic of YC6 cylinder with $L_x=10$ used in DMRG simulation. $\vec{y}$ is the direction with periodic boundary condition and $\vec{x}$ is the direction with open boundary condition.
}
\label{fig:yc6}
\end{figure}

\para
The extended Hubbard model of interest is 
\begin{equation}
     H =  -t\sum_{\left<ij\right>}\left(c_i^\dag c_j + H.c.\right)  
    + U\sum_{i} n_{i, \uparrow} n_{i, \downarrow} \\
     + V_1 \sum_{\left<ij\right>} n_i n_j,
\end{equation}
where $t, U, V_1$ represent hopping, on-site, and nearest neighboring Coulomb interaction strengths, respectively. The $\uparrow, \downarrow$ here should be interpreted as the valley degree of freedom $K, K'$. The phase space is thus spanned by two independent parameters, $U/t$ and $V_1/U$.   In prospect to experimental studies using TMD moir{\'e} systems, we focus on a parameter range that is accessible in hetero-bilayer TMD moir{\'e} experiments. We study the fixed filling of $\nu=1/3$ filling by density matrix renormalization group (DMRG) 
% The DMRG calculations are carried out on a 6-leg cylinder with the periodic boundary condition along one direction and open boundary condition along the other.
% Details are discussed in the Methods section.
% \textbf{Methods}
% \para We perform density matrix renormalization group (DMRG) 
\cite{White1992Phys.Rev.Lett.c} simulation, which is a powerful method for probing ground-state physics. We focus on the so-called YC6 geometry; an example of a YC6 cylinder is shown in Fig. \ref{fig:yc6}. Periodic boundary condition is imposed in the short direction ($\vec{y})$ and open boundary condition in the long direction ($\vec{x}$). Note that the choice of cylinder width $L_y=6$ is intended to be compatible with possible proposals of charge orders. Due to the low charge density $\nu=1/3$, we found narrow ladders, such as YC$3$ cylinders, are prone to finite size effects leading to 1D physics. We study cylinders with lengths $L_x=18, 24, 36$ and keep a large number of states (up to $m=10,000$) to get good convergence (see Appendix A for more details.) The DMRG calculation uses the ITensor library~\cite{Fishman2022SciPostPhys.Codebasesb}. 

\begin{figure}[]
  \includegraphics[width=0.8\columnwidth]{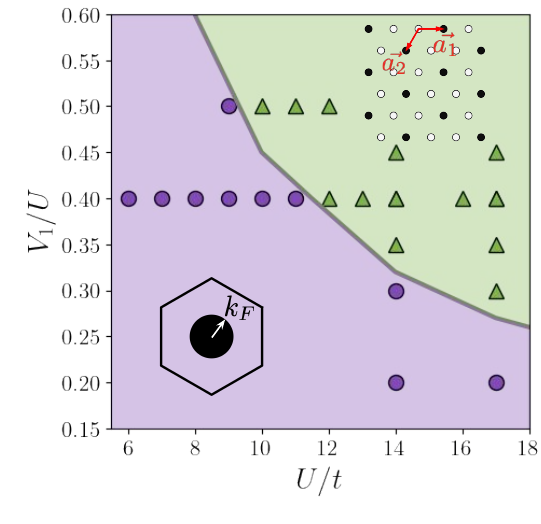}
  \caption{Phase diagram spanned by $U/t$ and $V_1/U$. The markers show where DMRG calculations are performed. The purple and green shading correspondingly represent the inferred extent of FL and GWC phases. The grey line is an indicator of the phase boundary. }
  \label{fig:phase_diagram}
\end{figure}

% \textbf{Results}
\para 
In the experimentally accessible phase space \footnote{The further range interaction strength can be estimated with a parallel capacitor model as used in Ref.\cite{Zhou2022Phys.Rev.Lett.}. See supplementary materials of \cite{Zhou2022Phys.Rev.Lett.} for details.} the weakly interacting limit is dominated by the Fermi liquid (FL),  as expected.  Surprisingly, the FL state is robust even in the limit of large on-site repulsion $U$, without the further range interaction $V_1$ (see the phase diagram in Fig. \ref{fig:phase_diagram}).  
This is in marked contrast to what has been found at so-called half-filling of $\nu=1$~\cite{szasz2020chiral, Zhou2022Phys.Rev.Lett., chen_chiral_2022,yang2023metalinsulator}, where on-site interaction alone can drive a metal-insulator transition into a chiral spin liquid phase without $V_1$~\cite{szasz2020chiral, Zhou2022Phys.Rev.Lett., chen_chiral_2022} although very small $V_1$ enhanced the chiral spin liquid behavior~\cite{Zhou2022Phys.Rev.Lett.}. 
% \EK{cite our previous paper.}  
The phase diagram illustrates the significance of the further range interaction at this low density, where electrons can be delocalized without much double occupancy.  Another striking feature of the phase diagram is its simplicity.  The insulating phase at above threshold values of $V_1/U$ is the so-called $\sqrt{3}\times\sqrt{3}$ phase predicted in the strong coupling limit Monte Carlo studies~\cite{Matty2022NatCommun} 
and observed in the flat band limit~\cite{Li2021Naturec}.
In the following discussion,  we track the evolution of various aspects of the many-electron state through the phase transition, focusing on a horizontal cut at fixed $V_1/U=0.4$ but increasing $U/t$.

\begin{figure}[]
  \includegraphics[width=0.9\columnwidth]{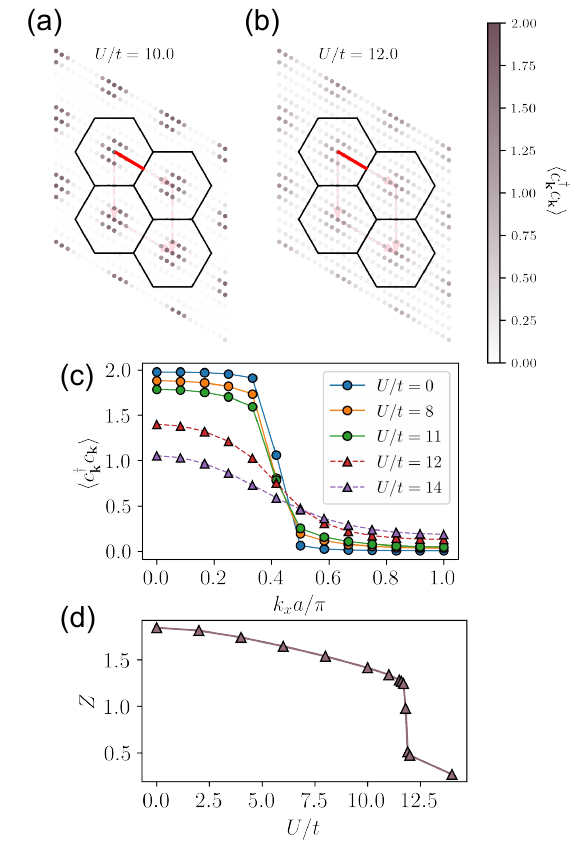}
  \caption{ Evolution of Fermi surface.
  (a)-(b) Momentum space occupancy at $U/t = 10.0$ and $12.0$ respectively, corresponding to FL and GWC phases. 
(c) Momentum space occupation $n(\mathbf{k}) = \langle c_{\mathbf{k}}^\dagger c_{\mathbf{k}}\rangle$ along a cut (indicated by red line in panel (a) and (b)) in the Brillouin zone. 
(d) Residue $Z$ at FS as a function of interaction strength $U/t$.  FS disappears around $U/t=11.8$ as evidenced by the abrupt drop of $Z$. 
}
\label{fig:fermi_surface}
\end{figure}
\para
First, we observe the loss of the coherent extended FL quasiparticles upon metal-insulator transition.
As shown in Fig. \ref{fig:fermi_surface}(a), the FL state at  $U/t=10.0$ shows momentum space occupation $\langle n_k\rangle = \langle c_k^\dag c_k\rangle$  defining the Fermi sea with the clearly defined Fermi surface. 
By contrast,  the momentum occupation is spread out in momentum space without a sign of Fermi surface in the insulating phase at $U/t=12.0$ (see Fig. \ref{fig:fermi_surface}(b)). 
The evolution of the momentum occupation along a cut in Brillouin zone (BZ) highlighted in red in Fig.\ref{fig:fermi_surface}(a) and (b),  tracked through the MIT in Fig.~\ref{fig:fermi_surface}(c), clearly shows the contrast between the FL state and the insulating state.  
We estimate the quasiparticle residue $Z$ as the difference between the momentum occupation at two momentum points bracketing the Fermi surface. 
As shown in Fig.\ref{fig:fermi_surface}(d),  the residue is continuously renormalized upon increasing the interaction strength  until the sudden drop near $U/t=11.8$.  This sudden drop signals discontinuous MIT through the loss of the sharp Fermi surface.  

\begin{figure}[]
  \includegraphics[width=1.0\columnwidth]{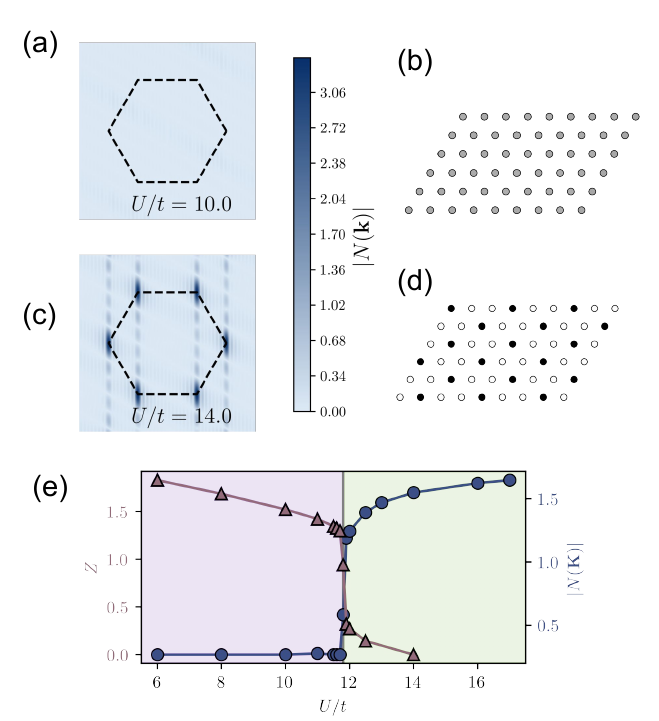}
  \caption{Charge order. Charge order parameter $N(\mathbf{k})$ at $U/t=10.0$ (a) and the corresponding real space charge order pattern (b). The same quantities at $U/t=14.0$ are shown in (c) and (d). The dashed lines in (a) and (c) represent the BZ of the triangular lattice. All calculation is done at $V_1/U=0.4$.   Peaks in $N(\mathbf{k})$ at $\mathbf{K}$ and $\mathbf{K'}$ points were observed in the charge-ordered phase. (e) Residue at Fermi surface $Z$ as a function of $U/t$ (blue curve). Fig. \ref{fig:fermi_surface}(d) is reproduced in the same plot (purple curve). The vertical grey line at $U/t\approx11.8$ marks the phase transition. The purple and green shaded regions represent FL and GWC phases, respectively.}
\label{fig:charge_order}
\end{figure}

\para 
The complete loss of Fermi surface 
is accompanied by the spontaneous breaking of lattice translational symmetry driven by strong further-range Coulomb interaction. 
The peak in the Fourier transform of the local density fluctuation $N(\mathbf{k})$ at $\mathbf{k}=\mathbf{K}$ for a particular $\mathbf{K}$ will signal the charge order with 
\begin{equation}
\label{eq:charge_order_parameter}
    N(\mathbf{k}) = \sum_{i} e^{-i \mathbf{k} \cdot \mathbf{r}_i} (\langle n_i \rangle - \nu),
\end{equation} 
where $\langle \cdot \rangle$ represents the number density expectation value with respect to the obtained ground state wave function and $\nu=1/3$ is the average density.
 For a clearer contrast between CO and FL, we remove the peak at zero wavevector by subtracting the average charge density in Eq. \ref{eq:charge_order_parameter}, since the zero wavevector peak exists even when the charge density is uniform.  %and the over-bar represents taking average over all lattice sites. In our case, $\overline{\langle n_i \rangle}=\nu=1/3$. 
Figures \ref{fig:charge_order}(a-d) unambiguously establishes that the insulating state has 
$\sqrt{3}\times\sqrt{3}$ order with modulation wavevectors
% $N(\mathbf{k}) \neq 0 $ for wavevetors $\mathbf{k}= (\frac{\pm 2\pi}{3a}, \frac{\pm 2\pi}{\sqrt{3}a}), (\frac{\pm 2\pi}{3a}, \frac{\mp 2\pi}{\sqrt{3}a}), (\frac{\pm 2\pi}{2\sqrt{3}a}, 0)$
% $\mathbf{k}= m \mathbf{k}_1 + n \mathbf{k}_2, m,n \in \mathbb{R}$, with 
$\mathbf{k}_1=(\frac{ 2\pi}{3a}, \frac{ 2\pi}{\sqrt{3}a}), \mathbf{k}_2 = (\frac{4\pi}{3a}, 0)$
($a$ is the triangular lattice constant), while the FL state has a uniform density.
%$N_\mathbf{k}\neq 0$ for finite $\mathbf{k}$ amounts to spontaneous breaking of lattice translational symmetry. As shown in Fig.\ref{fig:charge_order}(a), $N_\mathbf{k} = 0$ at $U/t=10.0$, which is within Fermi liquid phase. The charge distribution is uniformly spread in real space as in shown in Fig.\ref{fig:charge_order}(b). As we get into the charge ordered phase, we see emergence of peaks around BZ corners as shown in Fig.\ref{fig:charge_order}(c). The charge order we observe forms a triangular lattice with enlarged lattice constant, commonly referred to as the $\sqrt{3} \times \sqrt{3}$ charge order. A visualization of the charge order in real space is plotted in Fig.\ref{fig:charge_order}(d). 

%To study the transition into charge ordered phase, we look at the amplitude of the charge order parameter at the peaks, $|N_{K}|$, where $K$ represents one of the BZ corner, in Fig.\ref{fig:charge_order}(e). We see a wide region of $U/t$ where $N_K$ remains zero  until around $U/t=11.5$ where $N_K$ jumps to a finite value, indicating breaking of lattice transnational symmetry and formation of GWC. 

\para
Overlaying the plot of $N(\mathbf{K})$ with that of the quasiparticle residue $Z$ as shown in Fig.\ref{fig:charge_order}(e) clearly establishes that the abrupt 
 disappearance of Fermi surface and the sudden emergence of CO happens simultaneously at $U/t\approx11.8$ (Also see Appendix C). 
 A scan using a finer grid of $U/t$ near the transition point consistently showed the concurrence between the loss of the Fermi surface and the formation of the charge order. 
 Such concurrence rules out the possibility of an intermediate phase predicted by the HF mean field theory \cite{Pan2022Phys.Rev.Ba}. 
While it is hard to rule out a continuous transition where the charge order onsets steeply from our numerical results,  a direct continuous transition requires an exotic mechanism and fine-tuning.  
A recent study~\cite{Musser2022Phys.Rev.Bc} found that such a continuous transition is possible when the charge order in the insulating state occurs at longer wavelengths and the insulating state has a spinon Fermi surface. 
The fact that the charge order we observe is different from what was found to yield a continuous transition in Ref.\cite{Musser2022Phys.Rev.Bc} further supports the case that our numerical results indicate a first-order transition. 
 
\begin{figure}[]
\includegraphics[width=1.0\columnwidth]{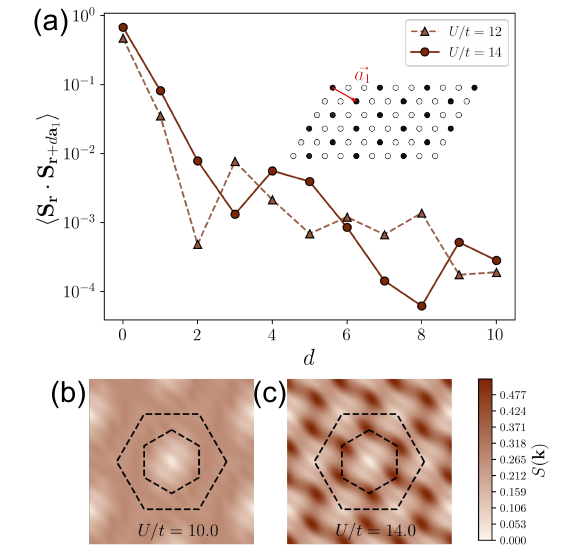}
  \caption{Spin correlation.  
(a) Spin correlation as a function of spin-spin separation $d$ along $\vec{a}_1$ direction for $U/t =12.0$ and $14.0$, both in GWC phase. The $\vec{a}_1$ direction is marked by red arrow in the inset.
(b)-(c) Spin structure factor $S(\mathbf{k})$ for $U/t=10.0$ (FL) and $U/t=14.0$ (GWC) respectively. The dashed lines represent the BZ of the original triangular lattice and reduced BZ caused by $\sqrt{3} \times \sqrt{3}$ charge order pattern.
}
\label{fig:spin_correlation}
\end{figure}

\para
We now turn to the spin (valley) channel. 
In Mott insulators expected at half-band filling of a single-band spinful lattice model,  nearest neighbor virtual hopping drives anti-ferromagnetic superexchange interaction between neighboring sites, leading to rich spin physics.  On the other hand, the WCs are expected to show ferromagnetism upon minute doping through a totally different so-called ``kinetic'' mechanism\cite{Nagaoka1966Phys.Rev., Davydova2023, Schlomer2023, Samajdar2023, Ciorciaro2023, Seifert2023}.   At $\nu=1$ (so-called ``half-filling''), anti-ferromagnetic interaction dominated over further range interaction in the charge channel, resulting in a spin density wave.
% with modulation $\mathbf{k}= (\frac{\pm 2\pi}{3a}, \frac{\pm 2\pi}{\sqrt{3}a}), (\frac{\pm 2\pi}{3a}, \frac{\mp 2\pi}{\sqrt{3}a}), (\frac{\pm 4\pi}{3a}, 0)$. 

\para
In the $\sqrt{3}\times\sqrt{3}$ charge-ordered GWC state, we found a long-range spin order driven by anti-ferromagnetic interactions. 
Fig.\ref{fig:spin_correlation}(a) shows the spin-spin correlation function 
$\langle \mathbf{S}_\mathbf{r} \cdot \mathbf{S}_{\mathbf{r}+d  \mathbf{a}_1}\rangle$ 
as a function of spacial separation $d$ between the two spins along $\mathbf{a}_1$ direction (see inset). The spin correlation shows power-law decay in the GWC, which manifests a quasi long-range spin order. 
To see the contrast between spin-ordered GWC phase and FL phase, we show the spin structure factor $S(\mathbf{k})= \frac{1}{N}\sum_{ij} e^{-i\mathbf{k} \cdot \mathbf{r}_{ij} }\langle \vec{S}_i \cdot \vec{S}_j \rangle$ for FL and GWC phases in 
Fig. \ref{fig:spin_correlation}(b) and (c) respectively. In the GWC phase, we see the emergence of peaks in $S(\mathbf{k})$ at the reduced Brillouin zone corners caused by the enlarged unit cell, which is consistent with the $120^{\circ}$ N{\'e}el order driven by antiferromagnetic interactions.  The power-law spin correlation decay and the features in the spin structure factor evidence the formation of a long-range anti-ferromagnetic spin order in the GWC phase. 
Comparing the energies between the fully polarized state ($S=N_e/2$) and the antiferromagnetic state ($S=0$), we find
$E(S=N_e/2)-E(S=0)=2.978(2)$ at $U/t=14.0, V_1/U=0.4$. The significantly lower energy for the antiferromagnetic state confirms that antiferromagnetic interactions dominate in the system. 

\para In summary, we studied the extended Hubbard model on the triangular lattice at $\nu=1/3$ using large-scale DMRG calculations, motivated by experiments on TMD moir{\'e} hetero-bilayers. We scanned the phase space parameterized by on-site repulsion strength $U/t$ and further-range interaction strength $V_1/U$ within the experimentally accessible parameter range. We found a direct
first-order transition from FL to GWC, with the charge order onsetting simultaneously with the vanishing of a sharp Fermi surface. The charge order takes the form of $\sqrt{3} \times \sqrt{3}$, which is consistent with the scanning tunneling microscopy observations in the strong coupling limit~\cite{Li2021Naturec} and the classical Monte Carlo simulations~\cite{Matty2022NatCommun}.
This observed charge order pattern is consistent with previous microscopic calculations using Hartree-Fock~\cite{Pan2022Phys.Rev.Ba} and momentum space exact diagonalization~\cite{Morales-Duran2022a}. In the spin sector, we observed a long-range spin order tendency driven by anti-ferromagnetic interactions in GWC.

\para 
Our findings harden the empirical observation that small changes in the TMD bilayers lead to widely different phenomena. We predict the AA-hetero bilayer, captured by extended Hubbard model studied in this paper, to go through a first order phase transition from FL into a GWC charge ordered phase with 120$^\circ$ AF spin order upon increasing the strength of interaction. However, closely related systems such as AB-hetero bilayer and AA homo-bilayer have shown strikingly different phenomena, namely Quantum Spin Hall effect in AB hetero-bilayer~\cite{zhang2021spin, li2021quantum, pan2022topological,tao2023giant} and Fractional Quantum Anomalous Hall in AA homo-bilayer~\cite{wu2019topological,cai2023signatures, zeng2023thermodynamic, park2023observation}, respectively. A key question for the field would be to understand what tips the balance between these different competing phenomena. In the continuum limit under magnetic field, the subject of competition between WC and FQH have been much studied~\cite{tsui1982two,zhu1993wigner,platzman1993quantum,ortiz1993new,pan2005transition,zhao2018crystallization,rosales2021competition}. Given the shallow depth of the moir{\'e} potential in TMD hetero-structures, exploring a continuum model approach such as ~\cite{yang2023metalinsulator}, which can capture both FQH and WC could be a productive future direction. 

\textit{Acknowledgements - } The authors thank Haining Pan, Jie Shan, Kin Fai Mak, Nicol{\'a}s Morales-Dur{\'a}n, Seth Musser, T. Senthil, Wenjin Zhao for helpful discussions. YZ thanks Miles Stoudenmire for help with using the ITensor library. YZ was supported in part by a New Frontier Grant from Cornell University’s College of Arts, Sciences and by the Cornell Center for Materials Research with funding from the NSF MRSEC program (DMR-1719875) and the National Science Foundation (Platform for the Accelerated Realization, Analysis, and Discovery of Interface Materials (PARADIM)) under Cooperative Agreement No. DMR-2039380. 
This research is funded in part by the Gordon and Betty Moore Foundation’s EPiQS Initiative, Grant GBMF10436 to E-AK. 
E-AK was supported by the Ewha Frontier 10-10 Research Grant.
DNS was supported by the U.S. Department of Energy, Office of Basic Energy Sciences under Grant No. DE-FG02-06ER46305. 

% \clearpage
\bibliography{bibliography}
\appendix
\onecolumngrid
\section{Computational details of DMRG simulations}
\begin{figure}[]
\includegraphics[width=1.0\columnwidth]{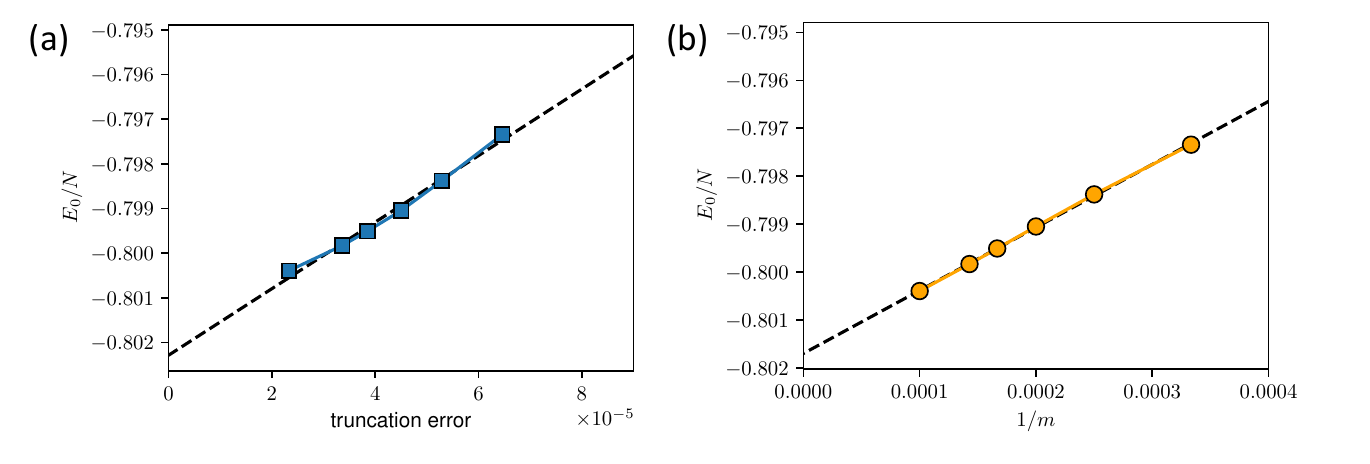}
  \caption{Energy convergence check for DMRG. Two ways to extrapolate true ground state energy per lattice site $E_0^*/N$ are presented. (a) Energy of ground state wavefunction obtained in DMRG per site, $E_0/N$, as a function of the truncation error (blue solid line). A linear line fit is shown as the black dashed line. Taking the limit truncation being zero, extrapolated $E_0^*/N=-0.80228642$. (b) $E_0/N$  as a function of inverse bond dimension $1/m$ (orange solid line). The black dashed line is a linear fit. Taking limit $m \rightarrow \infty$, or equivalently $1/m\rightarrow 0$, extrapolated $E_0^*/N=-0.80169494$. The data is for a $6 \times 24$ cylinder at $U/t=6.0, V_1/U=0.4$ corresponding to the FL phase. Since the FL phase is gapless, it is typically harder to converge than the gapped GWC phase.  }
\label{fig:dmrg_convergence}
\end{figure}

We use DMRG with $U(1)$ symmetry to study the extended Hubbard model, which preserves the total particle number in the system. We keep up to $10,000$ states (bond dimension $m=10000$) to get good convergence and the truncation error is on the order $\mathcal{O}(10^{-5})$. The phase space we scanned contains two phases, namely the FL and GWC phases. Since the FL is gapless, which is harder to converge than the gapped GWC phase, we here present the energy convergence check for a calculation in the FL phase in Fig.\ref{fig:dmrg_convergence}. 

\section{Real space spin correlation visualization}
\begin{figure}[]
\includegraphics[width=1.0\columnwidth]{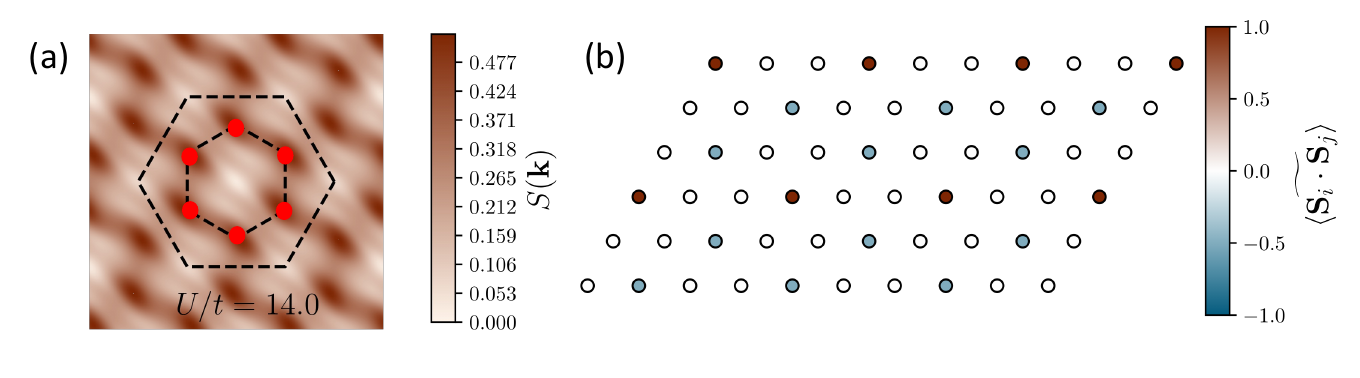}
  \caption{(a) The spin structure factor $S(\mathbf{k})$. The BZ of original triangular moire lattice and the mini BZ formed by $\sqrt{3} \times \sqrt{3}$ charge order are marked by black dashed lines. Red dots highlight the six-peak feature in $S(\mathbf{k})$. (b) Real space correlation function pattern corresponding to the six-peak feature in $S(\mathbf{k})$ (defined  in Eq.\ref{eq:six_peak}). }
\label{fig:spin_pattern}
\end{figure}

We calculate the spin structure factor $S(\mathbf{k})$ from spin correlation functions $\spincor$ obtained by measuring from DMRG ground state wavefunction, by a discrete Fourier transform: 
\begin{equation}
    S(\mathbf{k}) = \frac{1}{N}\sum_{\langle i j \rangle} \spincor e^{i \mathbf{k} \cdot \mathbf{r}_{ij}}.
\end{equation}
As shown in Fig.\ref{fig:spin_pattern}(a), we see features at corners of the reduced Brillouin zone, respecting $C_6$ symmetry. To see the real space pattern corresponding to such features, we simplify the spin structure factor to only keep the peaks at BZ corners and ignore all other fluctuations, which is highlighted by red dots in Fig.\ref{fig:spin_pattern}(a). In other words, we consider simplified spin structure factor
\begin{equation}
    \tilde{S}(\mathbf{k}) = \frac{1}{6} \sum_i \delta_{\mathbf{k}\mathbf{Q_i}}
\end{equation}
By discrete inverse Fourier transform, we can get the real space spin correlation 
\begin{equation}
\label{eq:six_peak}
    \widetilde{\spincor} = \sum_\mathbf{k} \tilde{S}(\mathbf{k}) e^{-i \mathbf{k} \cdot \mathbf{r}_{ij}} = \frac{1}{6} \sum_{i} e^{-i \mathbf{Q}_i \cdot \mathbf{r}_{ij} }.
\end{equation}
For better visualization, we choose $i=0$ as the ``center", such that $\widetilde{\spincor}$ captures the spin correlation between site $j$ and the chosen ``center" site $i=0$. We then can plotted $\widetilde{\spincor}$ on the triangular lattice as shown in Fig.\ref{fig:spin_pattern}(b). This real space spin correlation is clearly consistent with the famous $120^\circ$ N{\'e}el order. 
\section{Single particle Green's function}
\begin{figure}[]
\includegraphics[width=0.5\columnwidth]{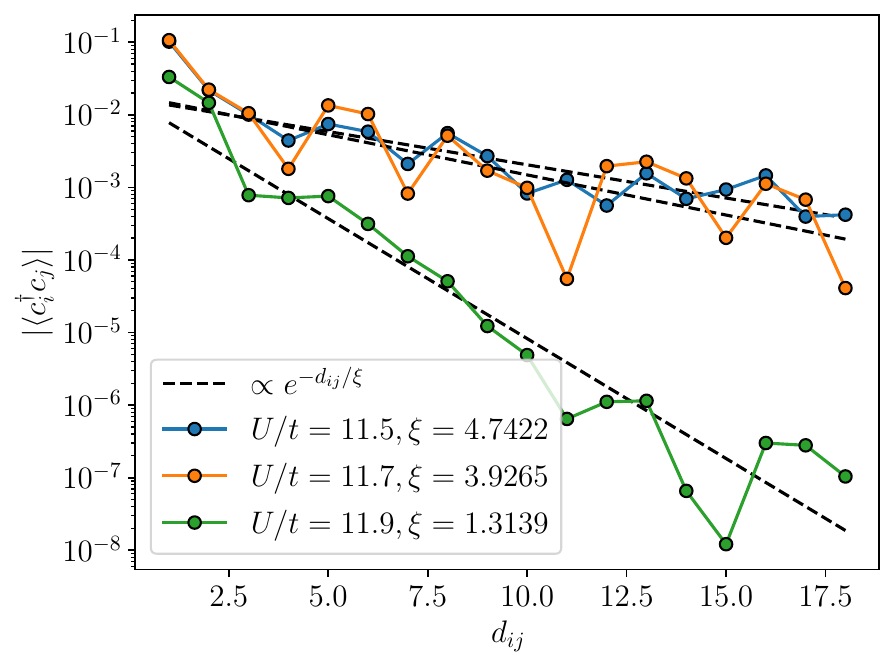}
  \caption{Magnitude of single particle Greens function $|\langle c_i^\dagger c_j\rangle|$ as a function of separation between sites $i,j$ denoted as $d_{ij}$ (solid lines with markers). The dashed lines indicate fitting to function of the form $\propto e^{-d_{ij}/\xi}$, which is used to extract correlation length $\xi$. }
\label{fig:single_particle_greens_function} 
\end{figure}
We show the magnitude of single particle Greens function $|\langle c_i^\dagger c_j\rangle|$ as a function of separation between sites $i,j$ in Fig.\ref{fig:single_particle_greens_function} for typical data points in both phases. By fitting to the function form $|\langle c_i^\dagger c_j\rangle|\propto e^{-d_{ij}/\xi}$, we extract the correlation length $\xi$, which are listed in the legend of Fig.\ref{fig:single_particle_greens_function}. Consistent with the sudden disappearance of FS and development of charge order at $U/t\sim 11.8$, we see the correlation length $\xi$ also show a clear decrease. In FL phase, the single particle Greens function decays slowly with increasing site separation distance $d_{ij}$. In contrast, in the GWC phase ($U/t=11.9$), the single particle Greens function decays very fast and has a very short correlation length.
\end{document}